# PROPOSING A CIPHERING PROTOCOL

Sarbajit Manna
Research Scholar
Vidyasagar University
Paschim Medinipur, Pin - 721102
sarbajitonline@gmail.com

Saurabh Dutta
Professor
Dr. B. C. Roy Engineering College
Jemua Road, Fuljhore, Durgapur - 713206
saurabh.dutta@bcrec.ac.in

## ABSTRACT
This paper describes a novel bit level stream cipher based symmetric key cryptographic technique. At first, sender and receiver agree upon a symmetric key. Then the symmetric key is formed using Greatest Common Divisor (G.C.D) of sum of even or odd bit positional weight[th] even or odd number and a symmetric key value is derived from the symmetric by the sender. The symmetric key value is XORed with the ASCII value of the plain text character to get the cipher text character. This is repeated for the entire plain text file. The symmetric key and the entire cipher text file is sent by the sender to the receiver. From the symmetric key, the receiver calculates the symmetric key value. The symmetric key value is XORed with the cipher text character to get back the plain text character. When this process is repeated for every cipher text character, then entire plain text file is formed by the decryption process. This technique has several advantages like formation of symmetric key dynamically, more emphasis on symmetric key value rather than the symmetric key, security and equal applicability for encryption and decryption of a large number of files of almost any type or extension.

## 1. INTRODUCTION
Symmetric Key Cryptography is a cryptographic technique where the sender and the receiver agree upon a common key. Then the sender uses the key to encrypt a message and the encrypted message is sent to the receiver over the network along with the key. The receiver decrypts the encrypted message using the key to get back the original message [11]-[13].

As the main focus of this technique is to enhance security, so instead of encrypting the plain text using the symmetric key, the plain text is encrypted using the symmetric key value which is generated from the symmetric key. During decryption, the plain text is re-generated from the cipher text using the symmetric key value which is derived from the symmetric key. The algorithm consists of following four major components [1],[2],[4].

### 1.1 Symmetric Key Generation
Plain Text ⟶ Symmetric Key

### 1.2 Symmetric Key Value Generation
Symmetric Key ⟶ Symmetric Key Value

### 1.3 Encryption Mechanism
Plain Text ⟷ Symmetric Key Value ⟶ Cipher Text (XOR)

### 1.4 Decryption Mechanism
Cipher Text ⟷ Symmetric Key Value ⟶ Plain Text (XOR)

Both the encryption and decryption technique is described in detail in section 2. An example implementation of the above mentioned technique is shown in tabular form in section 3 and the result and analysis are shown in section 4 which contains required analysis parameters and charts. Finally, conclusion is drawn in section 5.

## 2. PROPOSED ALGORITHM
An even number is a number which is evenly divisible by 2. An odd number is not evenly divisible by 2. It generates a remainder equal to 1 when it is divided by 2. 31 is an odd number because 31%2=1. 16 is an even number because 16%2=0, where % is the modulus operator [9]-[10].

Greatest Common Divisor (G.C.D) of two or more integers is the largest positive integer that divides the numbers without a remainder. The G.C.D of 54 and 24 is 6 as 6 is the largest positive integer that divides the numbers without a remainder [8].

Section 2.1 and section 2.2 respectively describe encryption and decryption mechanisms.

### 2.1 Encryption Mechanism
Following are the successive five steps of the encryption mechanism.

Binary stream of bits formation of the input file
↓
Symmetric Key generation from the Plain Text (PT)
↓
Symmetric Key value generation from the key
↓
Binary representation of the Symmetric Key value
↓
Cipher Text (CT) generation

Steps are described in section 2.1.1 to section 2.1.5.

*2.1.1 Binary stream of bits formation of the input file*
Entire plain text file is read from first to last one character at a time. Each character's ASCII value is converted into 8-bit binary representation.

Let, 'd' be the character read from the input file. The binary representation of 'd' (ASCII value 100) will be 01100100 which is represented in an array PT of 8-bits.

*2.1.2 Symmetric key generation from the Plain Text(PT)*
The technique uses a 32 bit symmetric key having 5 blocks of length 8 bits, 7 bits, 1 bit, 8 bits and 8 bits respectively. The 1[st] block represents sum of positional weights of odd numbered bit positions of the plain text ($S_{OP}$). The 2[nd] block represents sum of positional weights of even numbered bit positions of the plain text ($S_{EP}$). The 3rd block represents the condition $S_{OP} > S_{EP}$ which determines the numbers between which G.C.D will be performed. The 4th and the 5th block contain either $S_{OP}$ or $S_{EP}$ depending on the condition contained in 3rd block of the key. If the condition in the 3rd block is satisfied (represented by 1), then $S_{OP}$ is stored in 4th block of the key, otherwise $S_{EP}$ is stored here. If the condition in the 3rd block is not satisfied (represented by 0), then $S_{OP}$ is stored in 5th block of the key, otherwise $S_{EP}$ is stored here. All the entries in the key are stored in binary form.

The odd bit positions (OP) of a 8-bit number are : 1, 3, 5, 7 and the even bit positions (EP) are : 0, 2, 4, 6.

So, for plain text 'd', the OP positions are 1,3,5,7 and EP positions are 0,2,4,6. $S_{OP} = 0+0+32+0 = 32 = 00100000$ in binary, it is stored in 1st block of key and $S_{EP} = 0+4+0+64 = 68 = 1000100$ in binary, it is stored in 2nd block of key. As $S_{OP} < S_{EP}$, so 0 is stored in 3rd block of key. As the condition in the 3rd block of the key is not satisfied, so 01000100 ($S_{EP}$) and 00100000 ($S_{OP}$) is stored in 4th and 5th block of the key respectively.

### 2.1.3 *Symmetric Key value generation from the Plain Text(PT)*

Symmetric key value is the G.C.D of two numbers. These two numbers are calculated by using the following procedure. If the condition in the 3rd block is satisfied, then G.C.D is performed between $S_{OP}{}^{th}$ and $S_{EP}{}^{th}$ even terms starting from the ASCII value of the plain text character. If the condition in the 3rd block is not satisfied, then G.C.D is performed between $S_{EP}{}^{th}$ and $S_{OP}{}^{th}$ odd terms starting from the ASCII value of the plain text character.

So, for plain text 'd', G.C.D is performed between 68th and 32th ($S_{EP}{}^{th}$ and $S_{OP}{}^{th}$) odd terms from 100 (ASCII value of 'd'). The 68th and 32th odd terms from 100 are 235 and 163 respectively. G.C.D between 235 and 163 is 1. So the symmetric key value is 1.

### 2.1.4 *Binary representation of symmetric key value*

At first, the symmetric key value is represented in binary form. If the number of bits in the binary form is not a multiple of 8, then it is made multiple of 8 by inserting the remaining number of 0's to the left side of MSB, otherwise zero padding is not required.

So, for plain text 'd', $1_{10}$ is represented in equivalent binary form. The total number of bits in binary representation of 1 is 1, which is not a multiple of 8. It is made multiple of 8 by inserting seven 0's to the left side of MSB. $1_{10} = 1_2 = 00000001_2$ (By inserting seven 0's to the left of MSB to make it a multiple of 8).

### 2.1.5 *Cipher Text (CT) generation*

The symmetric key value is broken into a number of 8-bit blocks starting from MSB. Then the cipher text is generated by successively XORing the 8-bit blocks and the 8-bit binary representation of the plain text.

So, for plain text 'd', Cipher Text(CT) is generated by the following procedure where $\oplus$ represents XOR operation.

Symmetric Key Value = $00000001_2$

$00000001 \oplus PT_d(01100100)$

CT = 01100101

So, CT = $01100101_2 = 101_{10}$ = e, CT for $PT_d$ is e.

## 2.2 Decryption Mechanism
Following three steps represents the decryption mechanism.

Conversion of Cipher Text (CT) in binary form

Formation of symmetric key value from symmetric key

Plain Text (PT) generation

Steps are described in section 2.2.1 to section 2.2.3.

### 2.2.1 *Conversion of Cipher Text (CT) in binary form*
The encrypted file is received by the receiver and each character is converted into 8-bit binary representation from their ASCII values after reading it from the encrypted file.

So, Binary representation of cipher text 'e' is $01100101_2$ (ASCII value of e = $101_{10}$).

### 2.2.2 *Formation of symmetric key value from symmetric key*

As the symmetric key is known to the receiver, so the symmetric key value is generated by the receiver from the symmetric key. The symmetric key value is then represented in binary form. If the total number of bits in the binary representation of the symmetric key is not a multiple of 8, then it is made multiple of 8 by inserting remaining number of 0's to the left of MSB. Otherwise it is kept as it is.

For cipher text 'e', the receiver generates the symmetric key value $1_{10}$ and it is represented in binary form as $1_2$. The total number of bits in $1_2$ is 1, it is not a multiple of 8. So to make it multiple of 8, seven 0's (7) are padded to the left of MSB of $1_2$. So the binary representation of symmetric key value becomes $00000001_2$.

### 2.2.3 *Plain Text (PT) generation*
The symmetric key value is broken into a number of 8-bit blocks starting from MSB. Then the plain text is generated by successively XORing the 8-bit blocks and the 8-bit binary representation of the cipher text.

So, for cipher text 'e', Plain Text(PT) is generated by the following procedure where $\oplus$ represents XOR operation.

Symmetric Key Value = $00000001_2$

$00000001 \oplus CT_e(01100101)$

PT = 01100100

So, PT = $01100100_2 = 100_{10}$ = d. So $PT_d$ is successfully recovered from $CT_e$ by the receiver.

## 3. IMPLEMENTATION
There are two tables in this section. Table-1 and Table-2 respectively describes the detailed description of the encryption and decryption process of the proposed algorithm. A plain text file of size 1 KB (test.txt) is taken for encryption whose content is "do". Then the encrypted file ct_test.txt is formed from the plain text file test.txt. From ct_test.txt, the decrypted file pt_test.txt is formed by decryption. Then the contents of the plain text file test.txt and the decrypted file pt_test.txt are compared and it is seen that the contents are same.

**Table 1. Encryption Process**

| Plain Text (PT) | d (ASCII : 100) | o (ASCII : 111) |
|---|---|---|
| Binary equivalent of ASCII value | 01100100 | 01101111 |
| $S_{OP}$ | 32 (00100000) | 42 (00101010) |
| $S_{EP}$ | 68 (1000100) | 69 (1000101) |
| Condition : $S_{OP} > S_{EP}$ | 0 (Not Satisfied) | 0 (Not Satisfied) |
| $S_{OP} / S_{EP}$ | $S_{EP}$ = 68 (01000100) | $S_{EP}$ = 69 (01000101) |
| $S_{EP} / S_{OP}$ | $S_{OP}$ = 32 (00100000) | $S_{OP}$ = 42 (00101010) |
| $S_{OP}^{th}$ Even / $S_{EP}^{th}$ Odd from PT | $68^{th}$ Odd from 100 = 235 | $69^{th}$ Odd from 111 = 249 |
| $S_{EP}^{th}$ Even / $S_{OP}^{th}$ Odd from PT | $32^{th}$ Odd from 100 = 163 | $42^{th}$ Odd from 111 = 195 |
| Key Value (KV) | 1 (G.C.D between $68^{th}$ & $32^{th}$ odd number from 100) | 3 (G.C.D between $69^{th}$ & $42^{th}$ odd number from 111) |
| Binary equivalent of key value | 1 (No. of bits not a multiple of 8) | 11 (No. of bits not a multiple of 8) |
| Making key value multiple of 8 | Required 00000001 (Seven 0's inserted to the left) | Required 00000011 (Six 0's inserted to the left) |
| Cipher text (KV$\oplus$PT) in binary | 00000001 $\oplus$ 01100100 = 01100101 = $101_{10}$ | 00000011 $\oplus$ 01101111 = 01101100 = $108_{10}$ |
| Cipher Text (CT) | e (ASCII : 101) | l (ASCII : 108) |

**Table 2. Decryption Process**

| Cipher Text (CT) | e (ASCII : 101) | l (ASCII : 108) |
|---|---|---|
| Binary equivalent of ASCII value | 01100101 | 01101100 |
| Key Value (KV) generated from key | 1 (G.C.D between $68^{th}$ & $32^{th}$ odd number from 68+32=100) | 3 (G.C.D between $69^{th}$ & $42^{th}$ odd number from 69+42=111) |
| Binary equivalent of key value | 1 (No. of bits not a multiple of 8) | 11 (No. of bits not a multiple of 8) |
| Making key value multiple of 8 | Required 00000001 (Seven 0's inserted to the left) | Required 00000011 (Six 0's inserted to the left) |
| Plain text (KV$\oplus$CT) in binary | 00000001 $\oplus$ 01100101 = 01100100 = $100_{10}$ | 00000011 $\oplus$ 01101100 = 01101111 = $108_{10}$ |
| Plain Text (PT) | d (ASCII : 100) | e (ASCII : 108) |

## 4. RESULT AND ANALYSIS

Five different categories of files namely, .EXE, .DLL, .COM, .SYS, and .TXT are taken for result and analysis. Ten different files in terms of its name, size and content are taken for this section. Each sub-section consists of tables for each file category with information on source file size, target file size, encryption time, chi square value with the degree of freedom and avalanche percentage. The entire technique has been implemented using C language on a computer with Intel Pentium IV 2.40 GHz processor having 512 MB RAM.

To test the non-homogeneity between the source and the encrypted file, Pearson's chi-squared test has been performed. It means whether the observations onto encrypted files are in good agreement with a hypothetical distribution. In this case, the chi square distribution is being performed with (256-1)=255 degrees of freedom, 256 being the total number of classes of possible characters in the source as well as in the encrypted files. If the observed value of the statistic exceeds the tabulated value at a given level, the null hypothesis is rejected [3].

The "Pearson's chi-squared" or the "Goodness-of-fit chi-square" is defined by the following equation:

$$X^2 = \Sigma \{(f_0 - f_e)^2 / f_e\} \quad (1)$$

Here $f_e$ and $f_0$ respectively stand for the frequency of a character in the source file and that of the same character in the corresponding encrypted file. The chi-square values have been calculated on the basis of this formula for sample pairs of source and encrypted files [3].

The avalanche effect is a very important property of cryptographic algorithms. The avalanche effect is evident if, when an input is changed slightly, the output changes drastically. If a bit in the plain text is flipped or changed, then for an evident avalanche effect, almost half of the cipher text bits are changed in the cipher text. The small change can occur either in the plaintext or in the key so that it can cause a drastic change in the cipher text. In the proposed technique, the $3^{rd}$ bit of each plaintext is flipped. The key for each plaintext changes automatically with the flipping of bits in the plaintext. The avalanche percentage for each file is shown in following tables [5]-[7].

The result for .EXE, .DLL, .COM, .SYS, .TXT files is described in section 4.1, section 4.2, section 4.3, section 4.4 and section 4.5 respectively.

### 4.1 Result for .EXE files

The result for ten .EXE files is presented in Table 3. The file sizes vary from 5632 bytes to 21811 bytes. The encrypted file size, decrypted file size and the source file size is same. The encryption time varies from 83.21 seconds to 356.80 seconds. The chi square value lies in between 7560 to 30336 with the degree of freedom ranging from 208 to 255. The achieved avalanche is in the range from 13.35% to 18.86%.

**Table 3. Result For .EXE Files**

| Source/ Target File Size (Byte) | Encryption Time (S) | Chi Square Value | Degree of Freedom | Avalanche Achieved (%) |
|---|---|---|---|---|
| 12288 | 172.69 | 14342 | 208 | 13.35 |
| 11776 | 166.31 | 30336 | 252 | 16.65 |
| 11264 | 159.23 | 16268 | 251 | 17.24 |
| 8192 | 113.53 | 19010 | 251 | 16.74 |
| 5632 | 83.21 | 10390 | 231 | 15.67 |
| 7680 | 110.01 | 16506 | 245 | 16.36 |
| 18432 | 290.06 | 10163 | 255 | 17.20 |
| 15360 | 230.25 | 25955 | 253 | 16.45 |
| 21811 | 356.80 | 7560 | 255 | 18.86 |
| 12288 | 177.51 | 7690 | 226 | 13.50 |

The encryption time is graphically compared in Fig. 1 with the source file size for .EXE files. Each gray horizontal bar represents the source file size in KB and black horizontal bar represents the encryption time in second. It is seen that encryption time varies proportionally to the source file size.

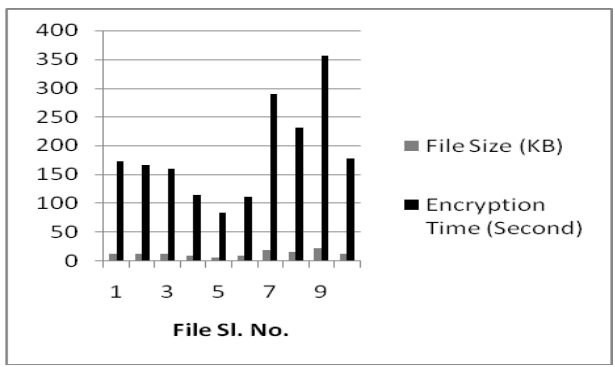

**Fig. 1: A comparison between source file size with encryption time for .EXE files**

### 4.2 Result for .DLL files

The result for ten .DLL files is presented in Table 4. The file sizes vary from 10752 bytes to 34816 bytes. The encrypted file size, decrypted file size and the source file size is same. The encryption time varies from 94.80 seconds to 517.18 seconds. The chi square value lies in between 1177 to 46103 with the degree of freedom ranging from 178 to 255. The achieved avalanche is in the range from 14.03% to 17.42%.

**Table 4. Result For .DLL Files**

| Source/ Target File Size (Byte) | Encryption Time (S) | Chi Square Value | Degree of Freedom | Avalanche Achieved (%) |
|---|---|---|---|---|
| 17408 | 218.11 | 6928 | 255 | 17.25 |
| 17408 | 94.80 | 12271 | 178 | 14.03 |
| 13312 | 160.60 | 4543 | 251 | 14.90 |
| 30208 | 441.43 | 20054 | 255 | 17.27 |
| 34816 | 517.18 | 9007 | 255 | 16.55 |
| 11264 | 130.28 | 30312 | 253 | 16.13 |
| 16896 | 208.28 | 1177 | 249 | 17.42 |
| 10752 | 128.09 | 22782 | 253 | 16.43 |
| 15872 | 194.66 | 23914 | 255 | 16.64 |
| 13312 | 154.67 | 46103 | 252 | 16.30 |

The encryption time is graphically compared in Fig. 2 with the source file size for .DLL files. Each gray horizontal bar represents the source file size in KB and black horizontal bar represents the encryption time in second. It is seen that encryption time varies proportionally to the source file size.

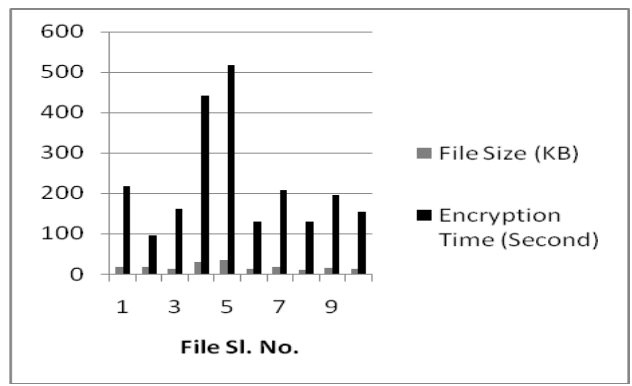

**Fig. 2: A comparison between source file size with encryption time for .DLL files**

### 4.3 Result for .COM files

The result for ten .COM files is presented in Table 5. The file sizes vary from 1126 bytes to 69836 bytes. The encrypted file size, decrypted file size and the source file size is same. The encryption time varies from 14.34 seconds to 1317.22 seconds. The chi square value lies in between -4486 to 19040 with the degree of freedom ranging from 181 to 255. The achieved avalanche is in the range from 16.28% to 18.07%.

**Table 5. Result For .COM Files**

| Source/ Target File Size (Byte) | Encryption Time (S) | Chi Square Value | Degree of Freedom | Avalanche Achieved (%) |
|---|---|---|---|---|
| 7680 | 108.92 | -4486 | 248 | 16.41 |
| 9216 | 130.12 | 6877 | 253 | 17.07 |
| 7168 | 100.79 | 13909 | 243 | 16.57 |
| 69836 | 1317.22 | 7216 | 255 | 18.07 |
| 29696 | 512.68 | 8470 | 254 | 17.51 |
| 26112 | 448.36 | 2457 | 255 | 17.75 |
| 19660 | 319.34 | 13376 | 255 | 17.70 |
| 7014 | 98.92 | 19040 | 255 | 16.28 |
| 14643 | 217.83 | 16001 | 255 | 17.22 |
| 1126 | 14.34 | -542 | 181 | 17.83 |

The encryption time is graphically compared in Fig. 3 with the source file size for .COM files. Each gray horizontal bar represents the source file size in KB and black horizontal bar represents the encryption time in second. It is seen that encryption time varies proportionally to the source file size.

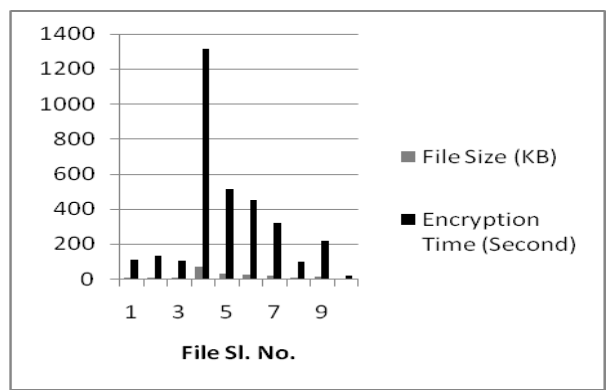

**Fig. 3: A comparison between source file size with encryption time for .COM files**

## 4.4 Result for .SYS files

The result for ten .SYS files is presented in Table 6. The file sizes vary from 4218 bytes to 16384 bytes. The encrypted file size, decrypted file size and the source file size is same. The encryption time varies from 49.21 seconds to 239.15 seconds. The chi square value lies in between -10749 to 54269 with the degree of freedom ranging from 229 to 255. The achieved avalanche is in the range from 15.79% to 17.78%.

**Table 6. Result For .SYS Files**

| Source/ Target File Size (Byte) | Encryption Time (S) | Chi Square Value | Degree of Freedom | Avalanche Achieved (%) |
|---|---|---|---|---|
| 10240 | 122.32 | -10749 | 254 | 17.12 |
| 11264 | 137.76 | -9917 | 253 | 16.78 |
| 15872 | 198.94 | 20031 | 255 | 17.38 |
| 10240 | 131.05 | 11376 | 250 | 17.05 |
| 10240 | 123.42 | 6663 | 254 | 17.15 |
| 9021 | 106.78 | 41964 | 253 | 17.78 |
| 4218 | 49.21 | 5489 | 229 | 15.99 |
| 16384 | 225.14 | 54269 | 244 | 16.09 |
| 4761 | 56.68 | 17431 | 247 | 17.26 |
| 17612 | 239.15 | 26135 | 254 | 15.79 |

The encryption time is graphically compared in Fig. 4 with the source file size for .SYS files. Each gray horizontal bar represents the source file size in KB and black horizontal bar represents the encryption time in second. It is seen that encryption time varies proportionally to the source file size.

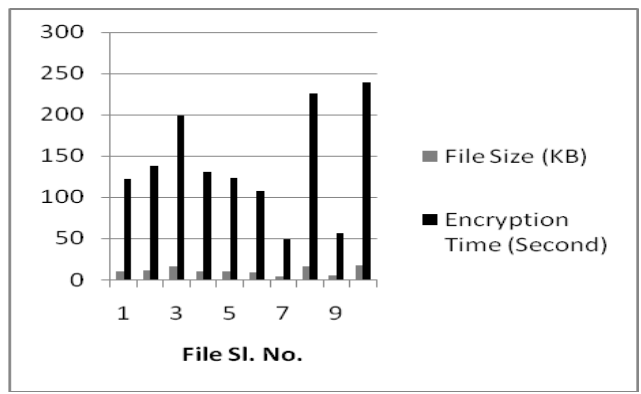

**Fig. 4: A comparison between source file size with encryption time for .SYS files**

## 4.5 Result for .TXT files

The result for ten .TXT files is presented in Table 7. The file sizes vary from 133 bytes to 32563 bytes. The encrypted file size, decrypted file size and the source file size is same. The encryption time varies from 0.61 seconds to 164.11 seconds. The chi square value lies in between -24290 to 23392 with the degree of freedom ranging from 42 to 98. The achieved avalanche is in the range from 11.26% to 15.97%.

**Table 7. Result For .TXT Files**

| Source/ Target File Size (Byte) | Encryption Time (S) | Chi Square Value | Degree of Freedom | Avalanche Achieved (%) |
|---|---|---|---|---|
| 1331 | 6.65 | 2126 | 80 | 15.92 |
| 5416 | 24.49 | 23392 | 87 | 14.77 |
| 6451 | 29.00 | 14327 | 87 | 11.26 |
| 133 | 0.61 | 230 | 42 | 14.66 |
| 829 | 3.79 | 1463 | 70 | 15.97 |
| 22937 | 104.02 | 20193 | 89 | 12.27 |
| 2201 | 11.04 | 5048 | 82 | 13.59 |
| 2498 | 11.31 | 2769 | 83 | 15.32 |
| 32563 | 164.11 | -24290 | 75 | 14.38 |
| 19046 | 85.84 | 11915 | 98 | 14.07 |

The encryption time is graphically compared in Fig. 5 with the source file size for .TXT files. Each gray horizontal bar represents the source file size in KB and black horizontal bar represents the encryption time in second. It is seen that encryption time varies proportionally to the source file size.

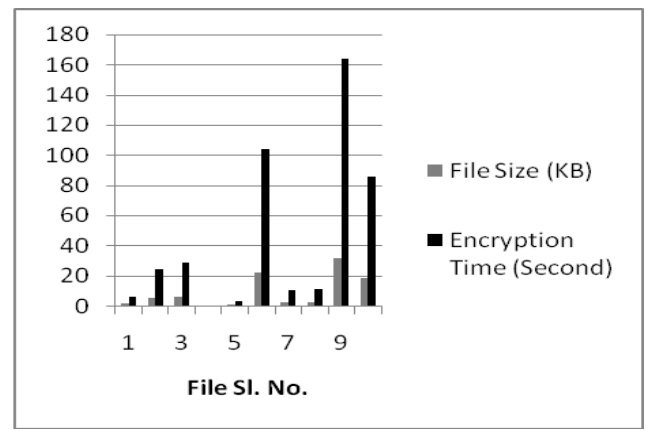

**Fig. 5: A comparison between source file size with encryption time for .TXT files**

## 5. CONCLUSION

The technique has been developed keeping two important factors in mind. The first among them is to make a dynamic key which is directly derived from the plain text character. The second factor is the security. To make the technique more secured, a symmetric key value is generated from the dynamic symmetric key which is operated on the plain text to get the cipher text. The symmetric key value is generated from the key using two stages. First, sum of even or odd bit positional weight[th] even or odd number starting from the ASCII value of plain text character is calculated. Then their G.C.D gives the symmetric key value. The key size is fixed. The key space is totally dependent on the source file size. The cipher text file size is same as the plain text file size. The technique is equally implemented in any popular high level language. It is also simple to implement. Extra memory space is not needed to store the encrypted and decrypted file. As the encryption and decryption has been done in bit level, the execution time is dependent on the source file size. The analysis parameter named achieved avalanche percentage is between 15 and 19 for most of the files. The avalanche percentage will be higher if more than one bit is flipped. The achieved avalanche percentage can also be calculated by flipping any one or more bits of the plain text, but also flipping any number of bits of the symmetric key or even both the plain text and the symmetric key. The cipher text can be formed not only using XOR and G.C.D operation but also any one or more reversible logical operations and L.CM (Least Common Multiple).

\* \* \* \* \* \*.